\documentclass[english]{article}
\usepackage{amssymb}
\usepackage{graphicx}
\usepackage{esint}

\makeatletter

\newcommand{\noun}[1]{\textsc{#1}}

\begin{document}

\title{Braids, 3-manifolds, elementary particles: number theory and symmetry
in particle physics }

\author{T. Asslmeyer-Maluga (German Aerospace Center, Berlin)}
\maketitle
\begin{abstract}
In this paper, we will describe a topological model for elementary
particles based on 3-manifolds. Here, we will use Thurston's geometrization
theorem to get a simple picture: fermions as hyperbolic knot complements
(a complement $C(K)=S^{3}\setminus(K\times D^{2})$ of a knot $K$
carrying a hyperbolic geometry) and bosons as torus bundles. In particular,
hyperbolic 3-manifolds have a close connection to number theory (Bloch
group, algebraic K-theory, quaternionic trace fields), which~will
be used in the description of fermions. Here, we choose the description
of 3-manifolds by branched covers. Every 3-manifold can be described
by a 3-fold branched cover of $S^{3}$ branched along a knot. In case
of knot complements, one will obtain a 3-fold branched cover of the
3-disk $D^{3}$ branched along a 3-braid or 3-braids describing fermions.
The whole approach will uncover new symmetries as induced by quantum
and discrete groups. Using the Drinfeld--Turaev quantization, we will
also construct a quantization so that quantum states correspond to
knots. Particle properties like the electric charge must be expressed
by topology, and we will obtain the right spectrum of possible values.
Finally, we will get a connection to recent models of Furey, Stoica
and Gresnigt using octonionic and quaternionic algebras with relations
to 3-braids (Bilson--Thompson model).

\end{abstract}
\tableofcontents

\section{Introduction}

{ General relativity (GR) deepens our view on space-time.
In parallel, the appearance of quantum field theory (QFT) gives us a different
view of particles, fields and the measurement process. One~approach for the unification of QFT and GR, to a quantum gravity,
starts with a proposal to quantize GR and its underlying structure,
space-time. Here, there is a unique opinion in the community about the relation
between geometry and quantum theory: the geometry as used in GR is
classical and should emerge from a quantum gravity in the usual limit that Planck's
constant tends to zero. Most theories went a step further and try
to get the spacetime directly from quantum theory. As~a~consequence, the used model of a smooth
manifold cannot be used to describe quantum gravity. However, currently, there is
no real sign for a discrete spacetime structure or higher dimensions in
current experiments \cite{FermiCollab2009}. Therefore, in this work, we conjecture
that the model of spacetime as a smooth 4-manifold can be also used in the quantum gravitational regime. As a consequence, one~has to find 
geometrical/topological representations for quantities in QFT 
(submanifolds for particles or fields, etc.)
as well in order to quantize GR.} In this paper, we will tackle this problem
to get a geometrical/topological description of the standard model
of elementary particle physics. Recently,~there~were efforts by Furey \cite{Furey2012,Furey2014,Furey2015,Furey2015_PhD}, Gresnigt
\cite{Gresnigt2018,Gresnigt2019_Group32,GresnigtGillard2019} and
Stoica \cite{Stoica2018} to use octonions and Clifford algebras
to get a coherent model to describe the particle generations in the
standard model. In the past, the stability of matter was related to
topology like in the approach of Lord Kelvin \cite{Kelvin1869} with
knotted aether vortices. The proposal to derive matter from space
was considered by Clifford as well by Einstein, Eddington, Schr{\"o}dinger
and Wheeler with only partial success (see \cite{MiThWh:73,Mielke1977}).
Giulini~\cite{Giulini09} discussed the status of geometrodynamics
in establishing particle properties like spin from spacetime by using
special solutions of general relativity. {The usage of knots and links to model particles, like~the electron, neutrinos, etc., was firstly observed by Jehle \cite{Jehle1981}. The phenomenological description of particle properties by using the quantum group $SU_q(n)$ is given in the work of Gavrilik \cite{Gavrilik2001}. Here,~the~(deformation) parameter $q$ of $SU_q(n)$ was linked with the flavor mixing angle (Cabibbo~angle). Furthermore, torus knots as given by 2-braids were associated with vector mesons (vector~quarkonia) of different flavors. Later, Finkelstein \cite{Finkelstein2009} used the representation of knots by quantum groups for his particle model.} Similar ideas are discussed
in the model of Bilson--Thompson \cite{Bilson--Thompson2005} in its
loop theoretic extension \cite{BTMarkSmolin2007}. 
Even for the Bilson--Thompson model, Gresnigt found
a link between Furey's approach and this model. However, some properties
of the Bilson--Thompson model remained mysterious, like~the definition
of the charge. Open is also the meaning of the braiding. If~there
is a connection between spacetime and matter, then one has to construct
the known fermions and bosons directly. Here,~it~seems that the main
problem is the determination of the underlying spacetime. In~this
paper, we will follow this way with an heuristic argument for the
spacetime to be the K3 surface in Section \ref{sec:Reconstructing-a-spacetime-K3}.
Then,~we~will analyze this spacetime by using branched covers to find
two suitable substructures, a knot complement representing the fermions
and a link complement to represent the bosons. Here,~we~profit from
ideas by Duston \cite{Duston2010,Duston2012,Duston2013} as well
from the work of Denicola, Marcolli and al-Yasry \cite{Marcolli_etal2010}.
As a byproduct, we also found interesting relations to octonions. The~representation of the knot and link complements by branched covers
gives the link to the original Bilson--Thompson model \cite{Bilson--Thompson2005}
but also to Gresnigt's work \cite{Gresnigt2018}. In Section \ref{sec:Electric-charge-quasimodularity},
we will discuss the electric charge and construct the corresponding
operator by using the underlying $U(1)$ gauge theory by using the
Hirzebruch defect. Finally, we will obtain the correct charge spectrum:
fermions carry the charges $0,\pm\frac{1}{3},\pm\frac{2}{3},\pm1$
in units of the unit charge $e$. Here, the factor $\frac{1}{3}$
is related to 4-dimensional topology (Hirzebruch signature theorem).
In Section~\ref{sec:Drinfeld--Turaev-quantization}, we will discuss
the independence of the particular braid from the particle. The~braid
is connected with the state of the particle as shown by using the
Drinfeld--Turaev quantization. Finally, we finish the paper with some
speculations about the number of generations, given by the number
of $S^{2}\times S^{2}$ parts in the spacetime (K3 surface), and a
global symmetry, the group $PGL(3,4)$ of $3\times3$ matrices of
the 4-element field $F_{4}$, induced from the K3 surface by using
umbral moonshine.

{
Before we start with the description of the model, we will discuss the key arguments and scope of the model. The model is based on a smooth spacetime that is described by a smooth 4-manifold. First,~it~is argued that this spacetime is the K3 surface where the evolution of the cosmos is submanifold. By using this model, the calculation of cosmic parameters (cosmological constant, inflation parameters, etc.) matches with the experimental results. For the following, we use the characterization of 4-manifold (i.e., the spacetime) by surfaces that are connected to complements of links and knots (3-manifolds with boundary). Interestingly, both 3-manifolds have a meaning by our previous work: knot complements are fermions and link complements are bosons. Here, we find a representation of both 3-manifolds by braids of three strands (3-braids), which is the connection to the work of Bilson--Thompson and Gresnigt. In case of the K3 surface as spacetime, the surfaces mentioned above are arranged with a strong connection to octonions, which is the link to Furey's work. To connect fermions, one needs a special class of link complements, so-called torus bundles, which can be interpreted as gauge fields. Interestingly, there are only three classes of torus bundles and we were able to connect them with the gauge groups $U(1),SU(2),SU(3)$. The main result of the paper is definition and interpretation of the electric charge. The electric charge is a topological invariant (Hirzebruch defect). For the fermions (knot complements), we obtained the spectrum $\{0,\pm1,\pm2,\pm3\}$ which agrees with the observation (normalized in $e/3$ units). The discussion about the quantization of the model and the number of generations closes the paper.

The model uses a classical spacetime. Quantum properties are not included in an obvious way. We obtained the correct charge spectrum, but we do not know the direct connection between knot complement and fermion (electron, neutrino, quark). There are ideas to use a quantization by a change of the smoothness structure, known as Smooth Quantum Gravity \cite{AsselmeyerMaluga2016}. In principle, the number of generations is connected with the spacetime. We got the minimal value of three generations, but it is only a lower bound. The model produces only the particles of the standard model. No supersymmetry or other extensions of the standard model can be derived by this model. Currently, we also have no idea to calculate the coupling constants and masses. It seems that these parameters are connected with the topological property of the spacetime.
}

\section{Preliminaries: Branched Coverings of 3- and 4-Manifolds}

According to Alexander (see \cite{Rol:76} for instance), every manifold
$M^{n}$ can be represented as $p$-fold branched covering $\pi:M^{n}\to S^{n}$
along an $n-2$ dimensional subcomplex $N^{n-2}\subset S^{n}$, the
branching set. In detail, the map $\pi$ is a covering except for
the branching set, i.e., for every point $b\in S^{n}\setminus N^{n-2}$
the map $\pi^{-1}(b)$ consists of $p$ points and the neighborhood
$U(b)$ is homeomorphic to one component $\pi^{-1}U(b)$. Then, $M^{n}\setminus\pi^{-1}(N^{n-2})\to S^{n}\setminus N^{n-2}$
is a $p-$fold covering. Usually, $N^{n-2}$ has the structure of a
simplicial subcomplex. The $p-$fold covering is completely determined
by the representation $\pi_{1}(S^{n}\setminus N^{n-2})\to S_{p}$
in the permutation group $S_{p}$ of $p$ symbols. Before going into
the details, we will look at some examples. First, there are no
branched coverings for 1-manifolds except for a trivial one. The first
interesting example is given by a compact 2-manifold, i.e., by a surface
of genus $g$. Here, the branching set consists of a finite number
of points ($0$-dimensional branching set). The branching set of 3-manifolds
is a one-dimensional, complex and we will later see that knots and links
are the appropriated structures. In case of 4-manifolds, one has a
two-dimensional complex as a branching set that was shown to be a surface.
These facts are easy to understand, but one parameter of a branched
covering is open: how many folds are minimally needed to represent
every manifold in a fixed dimension, or, how large is $p$ minimally
for a manifold of dimension $n$?

\subsection{As Warmup: Branched Coverings of 2-Manifolds}

\begin{figure}
\centering
\includegraphics[scale=0.3]{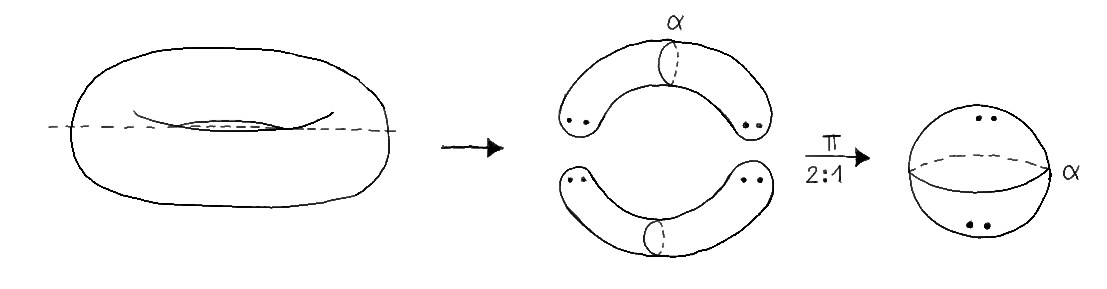}

\caption{2-fold covering of torus, $\alpha$ is the equator and the four-point
branching set. \label{fig:2-fold-covering-of-t2}}
\end{figure}

Let us start with the simplest case, the surface. By results of Riemann
and Hurwitz, every surface can be represented by a 2-fold covering
of $S^{2}$. As example, let us take the torus $T^{2}$ with the 2-fold
covering $T^{2}\to S^{2}$ branched along four points. The idea of the
construction is simple: choose a symmetry axis so that the genus $g$
surface can be generated by a rotation (see Figure \ref{fig:2-fold-covering-of-t2}).
 This axis meets the surface in 4g points which are the branching
points of the covering.

\subsection{Branched Coverings of 3-Manifolds}

For a 3-manifold, one has a one-dimensional, branching set and a result
of Alexander states that this branching set is a link with a finite
number of components. Later, Hirsch, Hilden and Montesinos \cite{Hilden1974,Hirsch1974,Montesinos1974}
obtained independently the result that every closed, compact 3-manifold
can be represented as a 3-fold branched covering of the 3-sphere branched
along a knot. As an example, consider the Poincare sphere $\Sigma(2,3,5)$
that can be represented by a 3-fold covering $\Sigma(2,3,5)\to S^{3}$
branched along the $(2,5)$ or $(5,2)$ torus knot. Now, let us consider
a closed, compact 3-manifold $N^{3}$ with a 3-fold branched covering
$N^{3}\to S^{3}$ branched along a knot $K$. It means that the map
$N^{3}\setminus K\to S^{3}\setminus K$ is a real $3:1$ map. Interestingly,
there is a diffeomorphism between $N^{3}\setminus K$ and $S^{3}\setminus K$
so that the $3:1$ covering map is now given by $S^{3}\setminus K\to S^{3}\setminus K$.
Furthermore, the 3-fold covering is completely determined by the map
$\pi_{1}(S^{3}\setminus K)\to S_{3}$, the representation of the fundamental
group into the permutation group $S_{3}$ of three letters. A simple
extension of the $S_{3}$ by considering the order of the permutations
gives the braid group $B_{3}$ of three strands. In principle, the
minimal number of folds $p=3$ is the root for the description of
particles by 3-braids, as shown later on.

\subsection{Branched Covering of 4-Manifolds}

\begin{figure}
\centering
\includegraphics[scale=0.3]{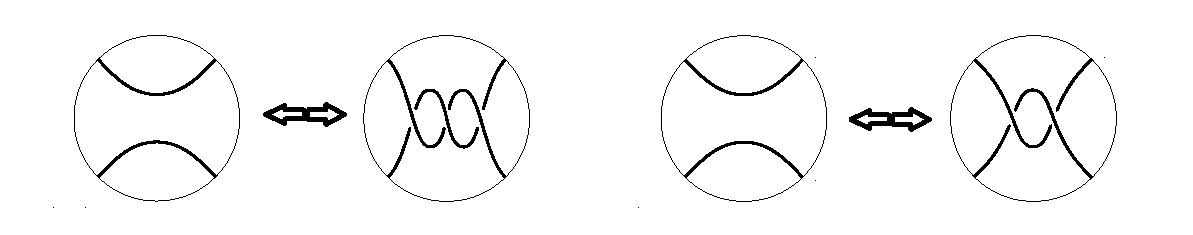}

\caption{Branching set changes, Left: Move 1, Right: Move 2 (so-called Montesino moves).
\label{fig:Branching-set-changes}}

\end{figure}

\begin{figure}
\centering
\includegraphics[scale=0.3]{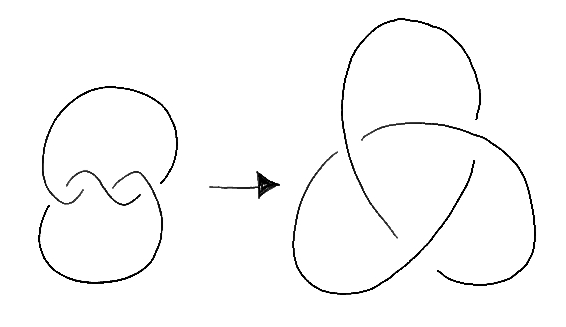}\includegraphics[scale=0.3]{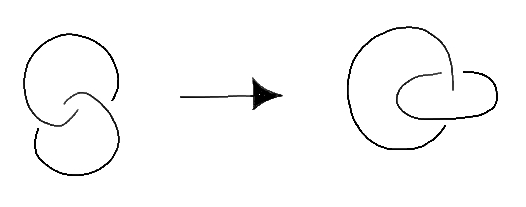}

\caption{Boundary of the branching set change---Left: Move 1 leading to the
Trefoil knot, Right: Move 2 leading to the Hopf link. \label{fig:Boundary-of-branching} }

\end{figure}

In a similar manner, one would expect that every 4-manifold $M^{4}$
can be represented as 4-fold branched covering of $S^{4}$ branched
along a surface. Piergallini \cite{Piergallini1995} was able to show
something similar, but the surface is only immersed and admits a finite
number of singularities (cusps and nodes). If one adds an additional
sheet, getting a 5-fold branched covering, then one can omit these
singularities (thus getting a locally flat embedded surface) \cite{PiergalliniIori2002}.
For a better understanding, we will discuss the way to this result
shortly. In \cite{Piergallini1995}, Piergallini considered the possible
transformations or changes of the branching set for a 3-manifold $N^{3}$,
i.e., the knot. Amazingly, he found two possible changes (see Figure
\ref{fig:Branching-set-changes}). 
All changes do not affect the underlying 3-manifold $N^{3}$, i.e.,
he found different knots and links representing the same 3-manifold
as a 3-fold branched cover. Then, he used this result to find a branched
covering for a 4-manifold. For that purpose, he introduced the concept
of an additional leaf or fold in the covering, i.e., if a 3-manifold
$N^{3}$ is represented by a 3-fold covering, then it is also represented
by a 4-fold covering. Then, he extended this 4-fold covering of $N^{3}$
to a 4-fold covering of $N^{3}\times[0,1]$ (i.e., a trivial cobordism).
At the same time, the knot $K_{1}$ as branching set of $N^{3}$at
one side of $N^{3}\times[0,1]$ (i.e., $N^{3}\times\left\{ 0\right\} $)
is related to the changed knot $K_{2}$ as branching set of $N^{3}$
on the other side of $N^{3}\times[0,1]$ (i.e., $N^{3}\times\left\{ 1\right\} $).
For the covering of $N^{3}\times[0,1]$, one will get a surface with
two boundaries, the disjoint union $K_{1}\sqcup K_{2}$ of the two
branching knots. {This procedure can be done for the two possible }changes \cite{PiergalliniZuddas2018}. For the first change,
Piergallini got the trefoil knot at the boundary, whereas, for the second
case, he obtained the Hopf link (see the Figures \ref{fig:Boundary-of-branching}).
 In dimension 4, one will get the cone over the trefoil, also known
as cusp, and the cone over the Hopf link, also known as node, as singularities
of the surface as a branching set of the 4-manifold. Expressed differently,
the trefoil knot is the link of the cusp singularity ${\displaystyle z^{2}+w^{3}}$;
the Hopf link (oriented correctly) is the link of the node singularity
${\displaystyle z^{2}+w^{2}}$. As explained in the previous subsection,
we have to consider the two fundamental groups $G_{P}=\pi_{1}(S^{3}\setminus\left\{ trefoil\right\} )$
and $G_{WW}=\pi_{1}(S^{3}\setminus\left\{ Hopf\right\} )$ for the
corresponding branched cover. Both groups are known and can be simply
calculated to be
\begin{equation}
G_{P}=\langle a,b\,|\, aba=bab\rangle=B_{3}\qquad G_{WW}=\langle a,b,c\,|\: ab^{-1}a^{-1}b\rangle=\mathbb{Z}\oplus\mathbb{Z},\label{eq:cusp-fold-singularities-branched-covering-4MF}
\end{equation}
which is a surprising result. From the point of 4-manifolds we will
get two possible 3-manifolds as boundary of the singularities: a 3-manifold
(knot complement) with one boundary and a 3-manifold (link complement)
with two boundaries. These two 3-manifolds can be simply interpreted:
the knot complement has one boundary component and can be seen as
a fermion (see also \cite{AsselmeyerBrans2015}) and the link complement
has two boundary components and can be interpreted as interaction
(see also \cite{AsselmeyerRose2012}).

\subsection{Branched Coverings of Knot Complements}

\begin{figure}
\centering
\includegraphics[scale=0.3]{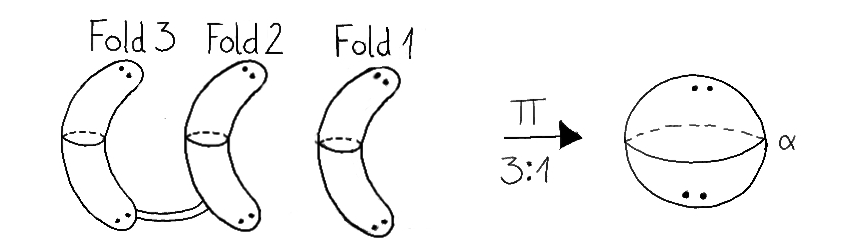}

\caption{3-fold covering torus.\label{fig:3-fold-covering-torus}}
\end{figure}

Now, we will describe the case of a 3-manifold (the knot complement)
with boundary a torus $T^{2}$. First, we will change the 2-fold
covering of the torus to a 3-fold covering by adding a trivial sheet.
For that purpose, we consider the 2-fold branched covering $T^{2}\to S^{2}$
and add a 2-sphere $T^{2}\#S^{2}\to S^{2}\#S^{2}$ which changes nothing
($S\#S^{2}$ is diffeomorphic to $S$ for every surface $S$). However,
at the same time, we will obtain a 3-fold covering (see Figure \ref{fig:3-fold-covering-torus}).

\begin{figure}
\centering
\includegraphics[scale=0.3]{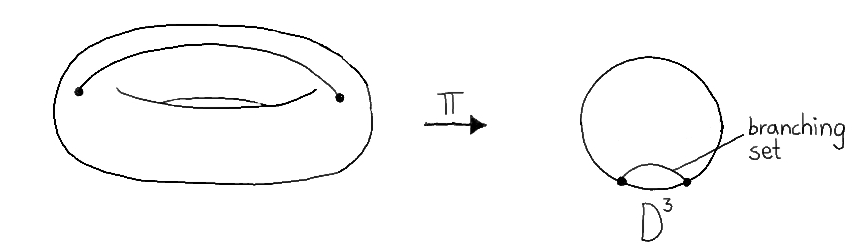}

\caption{Covering of the handle body.\label{fig:covering-of-handle}}
\end{figure}

\begin{figure}
\centering
\includegraphics[scale=0.3]{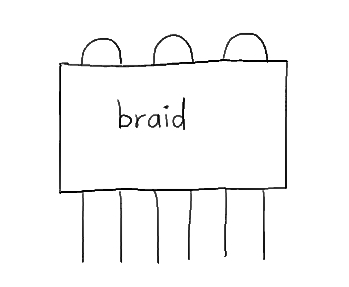}\includegraphics[scale=0.3]{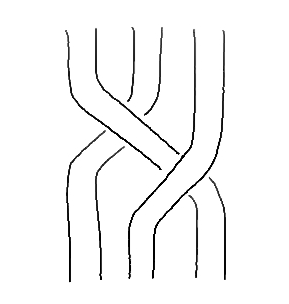}

\caption{\textbf{Left}: branching set of knot complement (6-plat), \textbf{Right}: 3-Braid as
6-Braid\label{fig:branching-set-braid}.}
\end{figure}

For a 3-manifold $\Sigma$ with boundary $T^{2}$, one has to consider
a branched covering $\Sigma\to S^{3}\setminus D^{3}$(3-sphere with
one puncture or $p$ punctures for $p$ boundary components). For
the construction of the covering, one needs another representation
of a 3-manifold, the Heegard decomposition. There, one considers two
handle bodies $H_{g},H'_{g}$ of genus $g$, i.e., the sum of $g$
copies of the solid torus $D^{2}\times S^{1}$. The gluing $H_{g}\cup_{\phi}H'_{g}$
of these handle bodies along the boundary using a diffeomorphism $\phi:\partial H_{g}\to\partial H'_{g}$(to
be precise: $\phi$ is an element of the mapping class group) produces
every compact, closed 3-manifold. For the 3-sphere, one obtains a
decomposition $H_{1}\cup H'_{1}$ using two solid tori where the meridian
of $H_{1}$ is mapped to the longitude of $H'_{1}$ and vice versa.
$H_{g}$ can be obtained by a branched covering $H_{g}\to D^{3}$
with a branching set $g+2$ arcs (see Figure \ref{fig:covering-of-handle}).
The diffeomorphism $\phi$ is represented by a braid connecting the
two handle bodies where the braid closes above and below to get a
link. In case of a 3-manifold with a boundary, the braid closes only
on one side and we obtain a braid again (or, more generally, a tangle),
see Figure \ref{fig:branching-set-braid}. 
 The underlying braid must be also a 3-braid but represented as a
6-braid.

\section{Reconstructing a Spacetime: The K3 Surface and Particle Physics\label{sec:Reconstructing-a-spacetime-K3}}

After so many years of experimental research, the standard model of
particle physics and of cosmology as well general relativity are confirmed
with high precision. There is no real contradicting result which shows
the necessity to introduce new physics. An exception may be cosmology
with the unknown components of dark matter and dark energy. However, this
situation is by no means satisfying. Both standard models have a bunch
of free parameters (19 parameters in particle physics, for instance).
If there is no sign for new physics, how did we get these parameters?
Here, we will argue that these parameters can be determined by topology.
However, at first glance, this idea seems hopeless. There are infinitely many
suitable topologies for the spacetime, seen as 4-manifold, and, for
the space, seen as 3-manifold. Here, we will go a different way. Why
not try to determine the space $\mathcal{M}$ of all possible spacetime-events?
Thus, let me start with a definition: let $\mathcal{M}$ be the space
of all possible spacetime events, i.e., the set of all spacetime events
carrying a manifold structure. In principle, $\mathcal{M}$ can be
identified with the spacetime. Then, a specific physical situation
is an embedding of a 3-manifold into $\mathcal{M}$, a dynamics is
an embedding of a cobordism between 3-manifolds into $\mathcal{M}$.
Here, we assume implicitly that everything can be geometrically/topologically
expressed as submanifolds (see \cite{AsselmeyerRose2012,AsselmeyerBrans2015}).
In the following, we will try to discuss this approach and how far
one can go. Some heuristic arguments are rather obvious: 
\begin{enumerate}
\item $\mathcal{M}$ is a smooth 4-manifold, 
\item any sequence of spacetime event has to converge to a spacetime event
and 
\item any loop (time-like or not) must be contracted. 
\end{enumerate}

A dynamics is a mapping of a spacetime event to a new spacetime event.
It is usually smooth (differential equations), motivating the first
argument. The second argument expresses the fact that any initial
spacetime event must converge to a final spacetime event, or the
limit of any sequence of spacetime events must converge to a spacetime
event. Then, $\mathcal{M}$ is a compact, smooth 4-manifold. The usual
spacetime is an open subset of $\mathcal{M}$. The third argument
above is motivated to neglect time-like loops. The spacetime is an
open subset of $\mathcal{M}$ or the spacetime is embedded in $\mathcal{M}$.
Now, consider a loop in the spacetime. By changing the embedding via
diffeomorphisms (this procedure is called isotopy), every loop is
contractable. Therefore, this argument implies that there is no time-like
loops (implying causality). Finally, $\mathcal{M}$ is a compact,
simply connected, smooth 4-manifold.

Now, we will restrict $\mathcal{M}$ in a manner so that we are able
to determine it. For the following, we implicitly assume that the equations
of general relativity are valid without any restrictions. Then, the
vacuum equations are equivalent to 
\[
R_{\mu\nu}=0,
\]
demanding Ricci-flatness. However, as shown in \cite{Bra:94a,Bra:94b}
and in recent years in \cite{AsselmeyerRose2012,AsselmeyerBrans2015,AsselmeyerMaluga2016},
the coupling to matter can be described by a change of the smoothness
structure. Therefore, the modification of the smoothness structure
will produce matter (or sources of gravity). However, at the same time,
we need a smoothness structure that can be interpreted as a vacuum
given by a Ricci-flat metric. Therefore, we will demand that 

\addtocounter{enumi}{3} 
\begin{enumerate}
\item $\mathcal{M}$ has {to admit a smoothness structure with Ricci-flat
metric representing the vacuum. }
\end{enumerate}

Interestingly, these four demands are restrictive enough to determine
the topology of $\mathcal{M}$. With the help of Yau's seminal work
\cite{Yau:1978}, we will obtain that   { $\mathcal{M}$ is homeomorphic to the K3 surface,} 
 using Yaus's work that there is only one compact, simply connected
Ricci-flat 4-manifold. However, it is known by the work of LeBrun \cite{Lebrun96}
that there are non-Ricci-flat smoothness structures. In the next step,
we will determine the smoothness structure of $\mathcal{M}$. For
that purpose, we will present some deep results in differential topology
of 4-manifolds: 
\begin{itemize}
\item there is a compact, contactable submanifold $A\subset\mathcal{M}$
(called Akbulut cork) so that cutting out $A$and reglue it (by an
involution) will produce a new smoothness structure, 
\item $\mathcal{M}$ splits topologically into 
\begin{equation}
|E_{8}\oplus E_{8}|\#\underbrace{\left(S^{2}\times S^{2}\right)\#\left(S^{2}\times S^{2}\right)\#\left(S^{2}\times S^{2}\right)}_{3\left(S^{2}\times S^{2}\right)}=2|E_{8}|\#3(S^{2}\times S^{2})\label{eq:splitting-K3}
\end{equation}
two copies of the $E_{8}$ manifold and three copies of $S^{2}\times S^{2}$
and 
\item the 3-sphere $S^{3}$ is a submanifold of $A$. 
\end{itemize}

In \cite{AsselmeyerKrol2012}, we already discussed this case. Interestingly,
there is always a topological 4-manifold for all combinations of $E_{8}$
and $S^{2}\times S^{2}$, but not all topological 4-manifolds are smooth
manifolds. Let us consider the 4-manifold that splits topologically
into $p$ copies of the $|E_{8}|$ manifold and $q$ copies of $S^{2}\times S^{2}$
or 
\[
p|E_{8}|\#q\left(S^{2}\times S^{2}\right)\:.
\]
Then, this 4-manifold is smoothable for every $q$ but $p=0$ and
the first combination for $p\not=0$ is the pair of numbers $p=2,q=3$
(which is the K3 surface). Any other combination ($p=2,q<3$ or every
$q$ and $p=1$) is forbidden as shown by Donaldson \cite{Don:83}.

Now, we consider the smooth K3 surface that is Ricci-flat, simply
connected, smooth. The main part in the following discussion will be the
use of the smoothness condition. As discussed above, the smoothness
structure is determined by the Akbulut cork $A$. Furthermore, as argued
above, the smoothness structure is strongly related to the appearance
of matter and this process is strongly connected to the evolution
of our cosmos. It is known as reheating after the inflationary phase.
Therefore, the Akbulut cork (including its embedding) should represent
the inflationary phase with reheating.

The Akbulut cork is built from a homology 3-sphere that will become
the boundary $\partial A$. The difference to a usual 3-sphere $S^{3}$
is given by the so-called fundamental group, the equivalence class
of closed loops up to deformation (homotopy) with concatenation as
group operation. In principle, one constructs a cobordism between
$S^{3}$ and the homology 3-sphere $\partial A$. All elements of
the fundamental group will be killed by adding appropriate disks.
At the end, one can add a 4-disk to get the full contractable cork
$A$. After this short discussion, we are able to identify the first
topological transition: if the cosmos starts as small 3-sphere (conjectural
of Planck size), then the space changes to $\partial A$, or 
\[
S^{3}\to\partial A.
\]
The topology of $\partial A$ depends strongly on the topology of
$\mathcal{M}$. In case of the K3 surface, $\partial A$ is known
to be a Brieskorn spheres, precisely the 3-manifold 
\[
\Sigma(2,5,7)=\left\{ x,y,z\in\mathbb{C}\,|\, x^{2}+y^{5}+z^{7}=0\,|x|^{2}+|y|^{2}+|z|^{2}=1\right\} \:.
\]
The embedding of the Akbulut cork is essential for the following results.
In \cite{AsselmeyerKrol2018a}, it was shown that the embedded cork
admits a hyperbolic geometry if the underlying K3 surface has an exotic
smoothness structure. This simple property has far-reaching consequences.
{Hyperbolic manifolds of dimension three or higher are rigid,
i.e., geometric properties like volume or curvature are topological
invariants (Mostow-Prasad rigidity). }If we assume that the cork $A$
represents the cosmic evolution, then geometric properties like the
curvature of $\partial A$ or the change of the size after the transition
$S^{3}\to\partial A$ are connected with topological properties of
the embedded cork $A$ and of the underlying K3 surface by using Mostow--Prasad
rigidity. This simple idea opens the door to explicit calculations.
In case of the transition $S^{3}\to\partial A=\Sigma(2,5,7)$, the
corresponding results can be found in \cite{AsselmeyerKrol2018a}.
If one assumes a Planck-size $(L_{P})$ 3-sphere at the Big Bang, then
the scale $a$ of $\Sigma(2,5,7)$ changes like 
\[
a=L_{P}\cdot\exp\left(\frac{3}{2\cdot CS(\Sigma(2,5,7))}\right)
\]
with the Chern--Simons invariant 
\[
CS(\Sigma(2,5,7))=\frac{9}{4\cdot(2\cdot5\cdot7)}=\frac{9}{280}
\]
and the Planck scale of order $10^{-34}m$ changes to $10^{-15}m$.
Obviously, this transition has an exponential or inflationary behavior.
Surprisingly, the number of e-folds can be explicitly calculated (see
\cite{AsselmeyerKrol2018b}) to be 
\begin{equation}
N=\frac{3}{2\cdot CS(\Sigma(2,5,7))}+ln(8\pi^{2})\approx51,\label{eq:e-fold-first}
\end{equation}
and we also obtain the energy and time scale of this transition (see
\cite{AsselmeyerKrol2018b,AsselmeyerKrol2018c}) 
\begin{equation}
E_{GUT}=\frac{E_{P}}{1+N+\frac{N^{2}}{2}+\frac{N^{3}}{6}}\approx10^{15}GeV\quad t=t_{P}\left(1+N+\frac{N^{2}}{2}+\frac{N^{3}}{6}\right)\approx10^{-39}s\label{eq:GUT-energy-scale}
\end{equation}
right at the conjectured scale of the Grand Unified Theory (GUT) ($E_{P},t_{P}$ Planck energy and
time, respectively). In our 
recent work \cite{AsselmeyerKrol2018c},
this transition was analyzed in a detailed manner. There, it was shown
that the transition can be described by a scalar field model which
conformally agrees (as shown in \cite{Whitt1984}) with the Starobinsky-$R^{2}$
theory \cite{Starobinski1980}. However, then, the dimension-less free
parameter $\alpha\cdot M_{P}^{-2}$ as well as spectral tilt ${\displaystyle n_{s}}$
and the tensor-scalar ratio $r$ can be determined to be 
\[
\alpha\cdot M_{P}^{-2}=1+N+\frac{N^{2}}{2}+\frac{N^{3}}{6}\approx10^{-5},\: n_{s}\approx0.961\,\: r\approx0.0046,
\]
using equation (\ref{eq:e-fold-first}), which is in good agreement with current
measurements. The embedding of the cork $A$ is based on the topological
structure of the K3 surface $\mathcal{M}$. As discussed above, $\mathcal{M}$
splits topologically into a 4-manifold $|E_{8}\oplus E_{8}|$ and
the sum of three copies of $S^{2}\times S^{2}$ (see \cite{GomSti:1999}).
In the topological splitting (\ref{eq:splitting-K3}), the 4-manifold
$|E_{8}\oplus E_{8}|$ has a boundary that is the sum of two Poincar{\'e}
spheres $P\#P$. Here, we used the fact that a smooth 4-manifold of
type $|E_{8}|$ must have a boundary (which is the Poincar{\'e} sphere
$P$); otherwise, it would contradict the Donaldson's theorem \cite{Don:83}.
Then, any closed version of $|E_{8}\oplus E_{8}|$ does not exist and
this fact is the reason for the existence of an exotic $\mathbb{R}^{4}$.
To express it differently, the neighborhood of the embedded cork lies
between the 3-manifold $\Sigma(2,5,7)$ (boundary of the cork) and
the sum of two Poincar{\'e} spheres $P\#P$. Therefore, we have two topological
transitions resulting from the embedding 
\begin{equation}
S^{3}\stackrel{cork}{\longrightarrow}\Sigma(2,5,7)\stackrel{gluing}{\longrightarrow}P\#P\,.\label{eq:two-transitions}
\end{equation}
The transition $\Sigma(2,5,7)\to P\#P$ has a different character
as discussed in \cite{AsselmeyerKrol2018a}. A direct consequence
is the appearance of a cosmological constant as a direct consequence
of the topological invariance of the curvature of a hyperbolic manifold.
With respect to the critical density, the final formula for normalized
cosmological constant, denoted by $\Omega_{\Lambda}$, reads 
\[
\Omega_{\Lambda}=\frac{c^{5}}{24\pi^{2}hGH_{0}^{2}}\cdot\exp\left(-\frac{3}{CS(\Sigma(2,5,7))}-\frac{3}{CS(P\#P)}-\frac{\chi(A)}{4}\right)\:.
\]
The Chern--Simons invariants $CS(P\#P)=\frac{1}{60},CS(\Sigma(2,5,7))=\frac{9}{280}$
and the Euler characteristics of the cork $\chi(A)=1$ together with
the Hubble constant (see \cite{PlanckCosmoParameters2013,PlanckCosmoParameters2015}
combined with \cite{HubbleTelescope2016}) 
\[
\left(H_{0}\right)_{Planck+Hubble}=69,2\,\frac{km}{s\cdot Mpc}
\]
gives the value 
\[
\Omega_{\Lambda}\approx0.7029,
\]
which is in excellent agreement with the measurements. The numerical
results above illustrate the power of the main idea to use topology
to fit the parameter in the standard model of cosmology. Interestingly,
these parameters are also important for particle physics. The existence
of two transitions (\ref{eq:two-transitions}) implies in the formalism
above the existence of two different energy scales, the GUT scale
of the first transition and a scale of Higgs mass order $126GeV$
for the second transition. These two scales are the right input for
the see-saw mechanism to generate a tiny neutrino mass (see \cite{AsselmeyerKrol2018b}).
Secondly, the formalism also provides a favor regarding the existence of a
right-handed neutrino. The energy scale of the two transitions $S^{3}\to\Sigma(2,5,7)\to P\#P$
can be expressed as a mass (via $Mc^{2}$), and we obtain 
\begin{equation}
M=\sqrt{\frac{4\hbar c}{G}}\left(\frac{\exp\left(-\frac{1}{2\cdot CS(P\#P)}\right)}{1+N+\frac{N^{2}}{2}+\frac{N^{3}}{6}}\right)\approx126.4GeV,\label{eq:Higgs-mass}
\end{equation}
which agrees with the mass of the Higgs boson (see \cite{AsselmeyerKrol2018b}).
Then, the Higgs boson can be expressed as the result of a topological
transition (see \cite{AsselmeyerKrol2014}). Now, we will use the two
energy scales to generate the neutrino mass. For that purpose, we start
with the non-diagonal mass matrix 
\[
\left(\begin{array}{cc}
0 & M\\
M & B
\end{array}\right)
\]
with two mass scales $B$ and $M$ fulfilling $M\ll B$. This matrix
has eigenvalues 
\[
\lambda_{1}\approx B\qquad\lambda_{2}\approx-\frac{M^{2}}{B}
\]
so that $\lambda_{1}$ is the mass of the right-handed neutrino, and
$\lambda_{2}$ represents the mass of the left-handed neutrino. Now,
we will use the two scales (\ref{eq:GUT-energy-scale}) and (\ref{eq:Higgs-mass})
\[
B\approx0.67\cdot10^{15}GeV,\quad M\approx126.4GeV,
\]
and we will obtain for the neutrino mass 
\[
m=\frac{M^{2}}{B}\approx0.024eV,
\]
which is in good agreement with the constraints from the PLANCK mission. 

These results seem to support our view that the K3 surface can be
the underlying spacetime (as seen as the set of all possible spacetime
events). The evolution of the cosmos is a suitable subset of this~
space.

\section{From K3 Surfaces to Octonions, 3-Braids and Particles}

The results of the previous section illustrated the power of the approach
and its relation to particle physics. In this section, we will discuss
the topological reasons and the relation to the models of Furey, Gresnigt
and Bilson--Thompson. In these models, 3-braids, octonions and quaternions
play a key role. Therefore, we have to understand how these structures
will naturally appear in the K3 surface.

\subsection{K3 Surfaces and Octonions\label{sub:K3-surfaces-and-octonions}}

The starting point for the description of any K3 surface is the topological
splitting (\ref{eq:splitting-K3})
\[
2|E_{8}|\#3(S^{2}\times S^{2}).
\]
The K3 surface is a closed, compact, simply connected 4-manifold.
According to Freedman \cite{Fre:82}, the topology is uniquely given
by the intersection form, a quadratic form on the second homology
characterizing the intersections of the generators as given by surfaces.
The K3 surface has the intersection form
\begin{equation}
Q_{K3}=E_{8}\oplus E_{8}\oplus(\oplus_{3}\left(\begin{array}{cc}
0 & 1\\
1 & 0
\end{array}\right)):=2E_{8}\oplus3H,\label{eq:intersection-K3}
\end{equation}
 with the the matrix $E_{8}$: 
\begin{equation}
E_{8}=\left[\begin{array}{cccccccc}
2 & 1 & 0 & 0 & 0 & 0 & 0 & 0\\
1 & 2 & 1 & 0 & 0 & 0 & 0 & 0\\
0 & 1 & 2 & 1 & 0 & 0 & 0 & 0\\
0 & 0 & 1 & 2 & 1 & 0 & 0 & 0\\
0 & 0 & 0 & 1 & 2 & 1 & 0 & 1\\
0 & 0 & 0 & 0 & 1 & 2 & 1 & 0\\
0 & 0 & 0 & 0 & 0 & 1 & 2 & 0\\
0 & 0 & 0 & 0 & 1 & 0 & 0 & 2
\end{array}\right].\label{eq:E8-matrix}
\end{equation}
This matrix is also the Cartan matrix of the Lie algebra $E_{8}$ and
here is where the connection to the octonions starts. For this purpose,
we have to deal with the root system of the Lie algebra $E_{8}$.
Consider a semi-simple Lie algebra $G$ and its Cartan subalgebra
$H=(H_{1},\ldots,H_{r})$, where $r$ is the rank of $G$. This subalgebra
is usually considered in the Cartan basis with the non-Hermitian generators
$E_{k}$ and their conjugates $E_{-k}$. $E_{k}$ is associated with
the root vector $r_{m}^{(k)}$ such that
\[
[H_{m},E_{\pm k}]=\pm r_{m}^{(k)}E_{\pm m}.
\]
When the rank $r$ is 1; 2; 4 or 8, we can combine the operators $H_{m}$
and the vectors $r_{m}^{(k)}$ (or eigenvalues) into elements of a
division algebra with imaginary units $e_{i}$: 
\[
H=H_{0}+e_{i}H_{i}\qquad r^{(k)}=r_{0}^{(k)}+e_{i}r_{i}^{(k)}
\]
 so that 
\[
[H,E_{\pm k}]=\pm r^{(k)}E_{\pm k}.
\]
For the Lie algebras of the groups $SU(2),O(4),O(8)$, one can construct
the real numbers $\mathbb{R}$, the complex numbers $\mathbb{C}$
and the quaternions $\mathbb{H}$, respectively. Interestingly, the
triality of the $O(8)$ group reflects the symmetry of the three quaternionic
units $e_{i}=-i\sigma_{i}$, where $\sigma_{i}$ are the Pauli matrices.
The case of $E_{8}$ was worked out by Coxeter \cite{Coxeter1946}
in connection with 8-dimensional regular solids and corresponds to
the octonions $\mathbb{O}$. There are 240 rational points on the
unit sphere $S^{7}$ represented by integer octonions that correspond
to the 240 roots of $E_{8}$. We first introduce octonionic imaginary
units $e_{\alpha}$ $(\alpha=1,\ldots,7)$ with the multiplication rule
\[
e_{\alpha}e_{\beta}=-\delta_{\alpha\beta}+\psi_{\alpha\beta\gamma}e_{\gamma},
\]
with $\psi_{\alpha\beta\gamma}$ as a third rank antisymmetric tensor
that is non-zero and equal to one for the index triples $123,246,435,367,651,572,714$.
Now, we define the special element
\[
h=\frac{1}{2}\left(e_{1}+e_{2}+e_{3}+e_{7}\right).
\]
Then, $1;e_{7};e_{2};e_{6},$ together with $h;eh;e_{2}h;e_{7}h$ correspond
to one possible set of principal positive roots. These elements are
also forming the Dynkin diagram of the root system of the $E_{8}$.
The matrix $E_{8}$ in equation (\ref{eq:E8-matrix}) above is the Cartan matrix
with entry $(i,j)$ defined by 
\[
2\frac{\langle r_{i},r_{j}\rangle}{\langle r_{i},r_{i}\rangle}
\]
the scalar products between the root vectors $r_{i}$. Then, the whole
approach showed that simple combinations of the octonionic imaginary
units are corresponding to generators of the second homology groups
for a 4-manifold having the matrix $E_{8}$ as an intersection form.
In case of the K3 surface, one has the intersection form containing
the matrix $E_{8}\oplus E_{8}$ which corresponds to two copies of
octonions $\mathbb{O}\times\mathbb{O}$. Here, there is a link to
the recent work \cite{GresnigtGillard2019} of complex sedions. Now,
every element of $\mathbb{O}$ is related to a surface (unique up
to homotopy). In the next section, we will present a connection of
these surfaces to spinors.

\subsection{From Immersed Surfaces in K3 Surfaces to Fermions and Knot Complements}

In the previous subsection, we described a relation between the octonions
and a system of eight intersecting surfaces in the K3 surface. In this
system of surfaces, every surface has two self-intersections (the
diagonal of the matrix (\ref{eq:E8-matrix})). Therefore, every surface
is not embedded but immersed in the K3 surface. For immersed surfaces,
there is a whole theory, called Weierstrass representation, with a
close connection between immersed surfaces and spinors. The following
discussion is borrowed from \cite{AsselmeyerRose2012}, and we will
present it here again for completeness. First, we start with the
immersion $I:\Sigma\to\mathbb{R}^{3}$ of a surface $\Sigma$ into
$\mathbb{R}^{3}$. This immersion $I$ can be defined by a spinor
$\varphi$ on $\Sigma$ fulfilling the Dirac equation
\begin{equation}
D\varphi=H\varphi,\label{eq:2D-Dirac}
\end{equation}
with $|\varphi|^{2}=1$ (or an arbitrary constant) (see Theorem 1
of \cite{Friedrich1998}). A spinor bundle over a surface splits into
two sub-bundles $S=S^{+}\oplus S^{-}$, representing spinors of different helicity, with the corresponding splitting of the spinor $\varphi$ in components
\[
\varphi=\left(\begin{array}{c}
\varphi^{+}\\
\varphi^{-}
\end{array}\right),
\]
and we have the Dirac equation
\[
D\varphi=\left(\begin{array}{cc}
0 & \partial_{z}\\
\partial_{\bar{z}} & 0
\end{array}\right)\left(\begin{array}{c}
\varphi^{+}\\
\varphi^{-}
\end{array}\right)=H\left(\begin{array}{c}
\varphi^{+}\\
\varphi^{-}
\end{array}\right)
\]
with respect to the coordinates $(z,\bar{z})$ on $\Sigma$. In dimension
3, the spinor bundle has the same fiber dimension as the spinor bundle
$S$ (but without a splitting $S=S^{+}\oplus S^{-}$into two sub-bundles).
Now, we define the extended spinor $\phi$ over the 3-torus $\Sigma\times[0,1]$
via the restriction $\phi|_{T^{2}}=\varphi$. The spinor $\phi$ is
constant along the normal vector $\partial_{N}\phi=0$ fulfilling
the three-dimensional Dirac equation
\begin{equation}
D^{3D}\phi=\left(\begin{array}{cc}
\partial_{N} & \partial_{z}\\
\partial_{\bar{z}} & -\partial_{N}
\end{array}\right)\phi=H\phi\label{eq:Dirac-equation-3D}
\end{equation}
induced from the Dirac equation (\ref{eq:2D-Dirac}) via restriction
and where $|\phi|^{2}=const.$ In this picture, we shift the description
from surfaces to 3-manifolds. The description above showed that the
essential\ information is contained in the surface, but fermions are
at least three-dimensional objects: fermions and bosons appear beginning
with dimension 3 (irreducible representation of the group $SO(3)$
as given by the lift to $SU(2)$). In dimension 2, we have anyons
with a spin of any rational number. However, how did we get the
corresponding 3-manifold representing the fermion?

To answer this question, we consider the branched covering of the K3
surface $M$. As explained above, it must be a 4-fold covering $M\to S^{4}$
branched along a surface with singularities of two types cusp and
fold. The cusp can be described as a cone over the trefoil knot, whereas
the fold is the cone over the Hopf link (see Figure 9 in \cite{PiergalliniZuddas2018}).
Now, we consider a 4-manifold with boundary, for instance by cutting
out a 4-disk $D^{4}$ form $M$ to get 4-manifold $M\setminus D^{4}$
with boundary $\partial(M\setminus D^{4})=S^{3}$, the 3-sphere. Then,
the branched covering of $M$ induced a branched covering of the boundary
$\partial M$, so that the branching set of $M$, a surface, induces
a branching set of $\partial M$, a knot or link. In our case, the
singularities of the surface (cusp and fold) given as cones over the
trefoil knot and Hopf link will correspond to the trefoil knot and
Hopf link in the 3-sphere. Then, the branched covering is given by
the mappings of the complements $S^{3}\setminus\left\{ trefoil\right\} $
and $S^{3}\setminus\left\{ Hopf-link\right\} $ to the permutation
group $S_{3}$. We see the appearance of these two complements as
a sign to use these structures as particles and interactions. The
complement of the knot is a 3-manifold with one boundary component.
In contrast, the complement of the link looks like a cylinder $T^{2}\times[0,1]$
which can connect two knot complements. Therefore, we have the conjecture
that knot complements are fermions and link complements are bosons.

\subsection{Fermions as Knot Complements}

In this section, we will discuss the topological reasons for the identification
of knot complements with fermions. In our paper \cite{AsselmeyerBrans2015},
we obtained a relation between an embedded 3-manifold and a spinor
in the spacetime. The main idea can be simply described by the following
line of argumentation. Let $\iota:\Sigma\hookrightarrow M$ be an
embedding of the 3-manifold $\Sigma$ into the 4-manifold $M$ with
the normal vector $\vec{N}$ so that a  small neighborhood $U_{\epsilon}$
of $\iota(\Sigma)\subset M$ looks like $U_{\epsilon}=\iota(\Sigma)\times[0,\epsilon]$.
Every 3-manifold admits a spin structure with a \noun{spin
bundle}, i.e., a principal $Spin(3)=SU(2)$ bundle (spin bundle) as
a lift of the frame bundle (principal $SO(3)$ bundle associated with
the tangent bundle). Furthermore, there is a (complex) vector bundle associated
with the spin bundle (by a representation of the spin group), called
\noun{spinor bundle} $S_{\Sigma}$. Now, we meet the usual definition in physics: a section in
the spinor bundle is called a spinor field (or a spinor). 
In general, the unitary representation of the spin group in $D$ dimensions
is $2^{[D/2]}$-dimensional. From the representational point of view,
a spinor in four dimensions is a pair of spinors in dimension 3. Therefore,
the spinor bundle $S_{M}$ of the 4-manifold splits into two sub-bundles
$S_{M}^{\pm}$ where one sub-bundle, say $S_{M}^{+},$ can be related
to the spinor bundle $S_{\Sigma}$ of the 3-manifold. Then, the spinor
bundles are related by $S_{\Sigma}=\iota^{*}S_{M}^{+}$ with the same
relation $\phi=\iota_{*}\Phi$ for the spinors ($\phi\in\Gamma(S_{\Sigma})$
and $\Phi\in\Gamma(S_{M}^{+})$). Let $\nabla_{X}^{M},\nabla_{X}^{\Sigma}$
be the covariant derivatives in the spinor bundles along a vector
field $X$ as section of the bundle $T\Sigma$. Then, we have the formula
\begin{equation}
\nabla_{X}^{M}(\Phi)=\nabla_{X}^{\Sigma}\phi-\frac{1}{2}(\nabla_{X}\vec{N})\cdot\vec{N}\cdot\phi\label{eq:covariant-derivative-immersion}
\end{equation}
with the embedding $\phi\mapsto\left(\begin{array}{c}
0\\
\phi
\end{array}\right)=\Phi$ of the spinor spaces from the relation $\phi=\iota_{*}\Phi$. Here,
we remark that, of course, there are two possible embeddings. For later
use, we will use the left-handed version. The expression $\nabla_{X}\vec{N}$
is the second fundamental form of the embedding where the trace $tr(\nabla_{X}\vec{N})=2H$
is related to the mean curvature $H$. Then, from (\ref{eq:covariant-derivative-immersion}),
one obtains the following relation between the corresponding Dirac
operators 
\begin{equation}
D^{M}\Phi=D^{\Sigma}\phi-H\phi\label{eq:relation-Dirac-3D-4D}
\end{equation}
with the Dirac operator $D^{\Sigma}$ on the 3-manifold $\Sigma$.
In \cite{AsselmeyerRose2012},
we extend the spinor representation of an immersed surface into the
3-space to the immersion of a 3-manifold into a 4-manifold according
to the work in \cite{Friedrich1998}. Then, the spinor $\phi$ defines
directly the embedding (via an integral representation) of the 3-manifold.
Then, the restricted spinor $\Phi|_{\Sigma}=\phi$ is parallel transported
along the normal vector and $\Phi$ is constant along the normal direction
(reflecting the product structure of $U_{\epsilon}$). However, then the
spinor $\Phi$ has to fulfill 
\begin{equation}
D^{M}\Phi=0\label{eq:Dirac-equation-4D}
\end{equation}
in $U_{\epsilon}$ i.e., $\Phi$ is a parallel spinor. Finally, we get
\begin{equation}
D^{\Sigma}\phi=H\phi\label{eq:Dirac3D-mean-curvature}
\end{equation}
with the extra condition $|\phi|^{2}=const.$ (see \cite{Friedrich1998}
for the explicit construction of the spinor with $|\phi|^{2}=const.$
from the restriction of $\Phi$). The idea of the paper \cite{AsselmeyerBrans2015}
was the usage of the Einstein--Hilbert action for a spacetime with
boundary $\Sigma$. The boundary term is the integral of the mean
curvature for the boundary; see \cite{Ashtekar08,Ashtekar08a}. Then
by the relation (\ref{eq:Dirac3D-mean-curvature}) we will obtain
\begin{equation}
\intop_{\Sigma}H\,\sqrt{h}\, d^{3}x=\intop_{\Sigma}\bar{\phi}\, D^{\Sigma}\phi\,\sqrt{h}d^{3}x\label{eq:relation-mean-curvature-action-to-dirac-action}
\end{equation}
using $|\phi|^{2}=const.$ As shown in \cite{AsselmeyerBrans2015},
the extension of the spinor $\phi$ to the 4-dimensional spinor $\Phi$
by using the embedding 
\begin{equation}
\Phi=\left(\begin{array}{c}
0\\
\phi
\end{array}\right)\label{eq:embedding-spinor-3D-4D}
\end{equation}
can be only seen as embedding, if (and only if) the 4-dimensional
Dirac equation 
\begin{equation}
D^{M}\Phi=0\label{eq:4D-Dirac-equation}
\end{equation}
on $M$ is fulfilled (using relation (\ref{eq:relation-Dirac-3D-4D})).
This Dirac equation is obtained by varying the action 
\begin{equation}
\delta\intop_{M}\bar{\Phi}D^{M}\Phi\sqrt{g}\: d^{4}x=0.\label{eq:4D-variation}
\end{equation}
In \cite{AsselmeyerBrans2015}, we went a step further and discussed
the topology of the 3-manifold leading to a fermion. On general grounds,
one can show that a fermion is given by a knot complement admitting
a hyperbolic structure. However, for hyperbolic manifolds (of dimension
greater than 2), one has the important property of Mostow rigidity
where geometric expressions like the volume are topological invariants.
This rigidity is a property which we should expect for fermions. The
usual matter is seen as dust matter (incompressible $p=0$). The scaling
behavior of the energy density $\rho$ for dust matter is determined
by the time-dependent scaling parameter $a$ to be $\rho\sim a^{-3}$.
Thus, if one represents matter by very small regions in the space equipped
with a geometric structure, then this scaling can be generated by an
invariance of these small regions with respect to a rescaling. Mostow
rigidity now singles out the hyperbolic geometry (and the hyperbolic
3-manifold as the corresponding small region) to have the correct
behavior. All other geometries allow a scaling at least along one
direction. Finally, \emph{Fermions are represented by hyperbolic knot
complements. }

\subsection{Torus Bundle as Gauge Fields}

Now, we have the following situation: two knot complements $C(K_{1})$
and $C(K_{2})$ can be connected by a so-called tube $T(K_{1},K_{2})$
along the boundary, a torus. This tube $T(K_{1},K_{2})$ can be described
by the complement of a link with two components defined by the knots
$K_{1},K_{2}$. In the simplest case, it is the 3-manifold $T^{2}\times[0,1]$.
The knot complements are fermions. Therefore, both knot complements
have to carry a hyperbolic structure, i.a. a space of constant negative
curvature. The frame bundle of a 3-manifold is always trivial, so
that we need a flat connection of this bundle to describe this space.
Let $Isom(\mathbb{H}^{3})=SO(3,1)$ be the isometry group of the three-dimensional
hyperbolic space. There are suitable subgroups $G_{1},G_{2}\subset Isom(\mathbb{H}^{3})$
so that (the interior of)$C(K_{1})$ is diffeomorphic to $Isom(\mathbb{H}^{3})/G_{1}$
(and similar with $C(K{}_{2})$). As usual, the space of all flat
$SO(3,1)$ connections of $C(K_{1})$ is the space of all representations
$\pi_{1}(C(K_{1}))\to SO(3,1)$, where $SO(3,1)$ acts in the adjoint
representation on this space (as gauge transformations). We note the
fact that every $SO(3,1)$ connections lifts uniquely to a $SL(2,\mathbb{C})$
connection. Now, near the boundary, we have a flat $SL(2,\mathbb{C})$
connection in $C(K_{1})$ which is connected to a flat $SL(2,\mathbb{C})$
connection in $C(K_{2})$ by $T(K_{1},K_{2})$. The action for a flat
connection $A$ with values in the Lie algebra $\mathfrak{g}$ of
the Lie group $G$ as a subgroup of the $SL(2,\mathbb{C})$ in a 3-manfold
$\Sigma$ (with vanishing curvature $F=DA=0$) is given by 
\[
\intop_{\Sigma}A\wedge F
\]
also known as background field model (BF model). By a small redefinition of the connection,
one can also choose the Chern--Simons action:
\[
CS(A,\Sigma)=\intop_{\Sigma}\left(A\wedge dA+\frac{2}{3}A\wedge A\wedge A\right).
\]
The variation of the Chern--Simons action $CS(A,\Sigma)$ gets flat
connections $DA=0$ as solutions. The flow of solutions $A(t)$ in
$T(K_{1},K_{2})\times[0,1]$ (parametrized by the variable $t$) between
the flat connection $A(0)$ in $T(K_{1},K_{2})\times\left\{ 0\right\} $
to the flat connection $A(1)$ in $T(K_{1},K_{2})\times\left\{ 1\right\} $
will be given by the gradient flow equation (see \cite{Flo:88} for
instance)
\begin{equation}
\frac{d}{dt}A(t)=\pm*F(A)=\pm*DA,\label{eq:gradient-flow}
\end{equation}
where the coordinate $t$ is normal to $T(K_{1},K_{2})$. Therefore,
we are able to introduce a connection $\tilde{A}$ in $T(K_{1},K_{2})\times[0,1]$
so that the covariant derivative in the $t$-direction agrees with $\partial/\partial t$.
Then, we have for the curvature $\tilde{F}=D\tilde{A}$ where the fourth
component is given by $\tilde{F}_{4\mu}=d\tilde{A}_{\mu}/dt$. Thus,
we will get the instanton equation with (anti-)self-dual curvature
\[
\tilde{F}=\pm*\tilde{F}\,.
\]
However, now we have to extend the Chern--Simons action (of the 3-manifold)
to the 4-manifold. It follows that
\[
CS(A,T(K_{1},K_{2})\times\left\{ 1\right\} )-CS(A,T(K_{1},K_{2})\times\left\{ 0\right\} )=\intop_{T(K_{1},K_{2})\times[0,1]}tr(\tilde{F}\wedge\tilde{F})
\]
i.e., we obtain the second Chern class and finally
\[
S_{EH}([0,1]\times T(K_{1},K_{2}))=\intop_{T(K_{1},K_{2})\times[0,1]}tr(\tilde{F}\wedge\tilde{F})=\pm\intop_{T(K_{1},K_{2})\times[0,1]}tr(\ \tilde{F}\wedge*\tilde{F})\,
\]
i.e., the action of the gauge field. The whole procedure remains true
for an extension, i.e.,
\begin{equation}
S_{EH}(\mathbb{R}\times T(K_{1},K_{2}))=\pm\intop_{T(K_{1},K_{2})\times\mathbb{R}}tr(\ \tilde{F}\wedge*\tilde{F})\,.\label{eq:gauge-action-tubes}
\end{equation}
The gauge field action (\ref{eq:gauge-action-tubes}) is only defined
along the tubes $T(K_{1},K_{2})$. For the extension of the action
to the whole 4-manifold $M$, we need some non-trivial facts from
the theory of 3-manifolds. We presented the ideas in \cite{AsselmeyerRose2012}.
Finally, we obtain the gauge field action
\begin{equation}
\intop_{M}tr(\tilde{F}\wedge*\tilde{F}).\label{eq:gauge-field-action}
\end{equation}
Now, we will discuss the possible gauge group. Again, for completeness,
we will present the argumentation in \cite{AsselmeyerRose2012} again.
The gauge field in the action (\ref{eq:gauge-field-action}) has values
in the Lie algebra of the maximal compact subgroup $SU(2)$ of $SL(2,\mathbb{C})$.
However, in the derivation of the action, we used the connecting tube $T(K_{1},K_{2})$
between two tori, which is a cobordism. This cobordism $T(K_{1},K_{2})$
is also known as torus bundle (see \cite{Calegari2007} Theorem 1.15),
which can be always decomposed into three elementary pieces---\emph{finite
order}, {Dehn twist} and the so-called Anosov map (named after the russian mathematician Dmitri Anosov).
 
The idea of this construction is very simple: one starts with two
trivial cobordisms $T^{2}\times[0,1]$ and glues them together by using
a diffeomorphism $g:T^{2}\to T^{2}$, which we call \emph{gluing diffeomorphism}.
From the geometrical point of view, we have to distinguish between
three different types of torus bundles. The three types of torus bundles
are distinguished by the splitting of the tangent bundle: 
\begin{itemize}
\item finite order (orders $2,3,4,6$): the tangent bundle is three-dimensional, 
\item Dehn-twist (left/right twist): the tangent bundle is a sum of a two-dimensional
and a one-dimensional bundle,
\item Anosov: the tangent bundle is a sum of three one-dimensional bundles.
\end{itemize}

Following Thurston's geometrization program (see \cite{Thu:97}),
these three torus bundles are admitting a geometric structure, i.e.,
it has a metric of constant curvature. Apart from this geometric properties,
all torus bundles are determined by the gluing diffeomorphism $g:T^{2}\to T^{2}$, {which
also determines the fundamental} 
group of the torus bundle. Therefore,
this gluing diffeomorphism also has an influence on the structure of
the diffeomorphism group of the torus bundle, which will be discussed
now. From the physical point of view, we have two types of diffeomorphisms:
local and global. Any coordinate transformation can be described by
an infinitesimal or local diffeomorphism (coordinate transformation).
In contrast, there are global diffeomorphisms like an orientation reversing
diffeomorphism. Two diffeomorphisms not connected via a sequence of
local diffeomorphism are part of different connecting components of
the diffeomorphism group, i.e., the set of isotopy classes $\pi_{0}(Diff(M))$
(also called the mapping class group). Isotopy classes are important in order
to understand the configuration space topology of general relativity
(see Giulini \cite{Giulini94}). In principle, the state space in geometrodynamics is the set of all isotopy classes, where every class represents one physical situation, or isotopy classes label two different physical situations. By definition, the two 3-manifolds in different isotopy classes cannot be
connected by a sequence of local diffeomorphisms (local coordinate
transformations). Again, these two different isotopy classes represent
two different physical situations; see \cite{Giulini09} for the relation
of isotopy classes to particle properties like spin. In case of the
torus bundle, we consider the isotopy classes $\pi_{0}(Diff(M,\partial M))$
relative to the boundary represented by the automorphisms of the fundamental
group. Using the geometrization program, we obtain a relation between
the isotopy classes $\pi_{0}(Diff(M,\partial M))$ and the isometry
classes (connecting components of the isometry group) with respect
to the geometric structure of the torus bundle (see, for instance, \cite{KalliongisMcCullough1996,HatcherMcCullough1997}).
Then, the isotopy classes of the torus bundles are given by 
\begin{itemize}
\item finite order: 2 isotopy classes (= no/even twist or odd twist),
\item Dehn-twist: 2 isotopy classes (= left or right Dehn twists),
\item Anosov: 8 isotopy classes (= all possible orientations of the three
line bundles forming the tangent bundle).
\end{itemize}

From the geometrical point of view, we can rearrange the scheme above:
\begin{itemize}
\item torus bundle with no/even twists: one isotopy class,
\item torus bundle with twist (Dehn twist or odd finite twist): three isotopy
classes,
\item torus bundle with Anosov map: eight isotopy classes.
\end{itemize}

This information creates a starting point for the discussion on how to derive
the gauge group. Given a Lie group $G$ with Lie algebra $\mathfrak{g}$, the rank of $\mathfrak{g}$ is the dimension of the maximal abelian
subalgebra, also called Cartan algebra (see above for a definition).
It is the same as the dimension of the maximal torus $T^{n}\subset G$.
The curvature $F$ of the gauge field takes values in the adjoint
representation of the Lie algebra and the action $tr(F\wedge*F)$
forms an element of the Cartan subalgebra (the Casimir operator).
However, each isotopy class contributes to the action and therefore we
have to take the sum over all possible isotopy classes. Let $t_{a}$
be the generator in the adjoint representation; then, we obtain for
the Lie algebra part of the action $tr(F\wedge*F)$
\begin{itemize}
\item torus bundle with no twists: one isotopy class with $t^{2}$,
\item torus bundle with twist: three isotopy classes with $t_{1}^{2}+t_{2}^{2}+t_{3}^{2}$,
\item torus bundle with Anosov map: eight isotopy classes with $\sum_{a=1}^{8}t_{a}^{2}$.
\end{itemize}

The Lie algebra with one generator $t$ corresponds uniquely to the
Lie group $U(1)$ where the three generators $t_{1},t_{2},t_{3}$ form
the Lie algebra of the $SU(2)$ group. Then, the last case with eight generators
$t_{a}$ has to correspond to the Lie algebra of the $SU(3)$ group.
We remark the similarity with an idea from brane theory: $n$ parallel
branes (each decorated with an $U(1)$ theory) are described by an
$U(n)$ gauge theory (see \cite{GiveonKutasov99}). Finally, we obtain
the maximal group $U(1)\times SU(2)\times SU(3)$ as a gauge group for
all possible torus bundles (in the model: connecting tubes between
the solid tori).

At the end, we will speculate about the identification of the isotopy
classes for the torus bundle with the vector bosons in the gauge field
theory. Obviously, the isotopy class of the torus bundle with no twist
must be the photon. Then, the isotopy class of the other bundle of
finite order should be identified with the $Z^{0}$ boson and the
two isotopy classes of the Dehn twist bundles are the $W^{\pm}$ bosons.
Here, we remark that this identification is consistent with the definition of the charge lateron.
Furthermore, we remark that this scheme contains automatically the mixing between
the photon and the $Z^{0}$ boson (the corresponding torus bundle
are both of finite order). The isotopy classes of the Anosov map bundle
have to correspond to the eight gluons. Later, we expect that these ideas will lead to an additional relation for the scattering amplitudes induced by the geometry of the torus bundles.

\subsection{Fermions, Bosons and 3-Braids}

In the previous subsection, we identified the hyperbolic knot complement
with a fermion and the torus bundle between them as gauge bosons.
The first natural question is then: which knot is related to a fermion
like the electron? However, this question is meaningless in our approach.
The knot/link complements are induced from the singularities of the
branching set of the 4-fold branched covering of the K3 surface. However,
these singularities are another expression of the change of the branching
set of the 3-manifolds. With other words: the branching set, a knot
or link, of a 3-manifold is not unique. There are transformations
of the branching sets representing the same 3-manifold. Therefore,
these complements are not uniquely connected to particles/fermions.
Later, we will see that the knots represent mainly the state. However, what
are the invariant properties? For that purpose, we will study the branched
covering of the knot/link complement to understand the invariant properties.

\begin{figure}
\centering
\includegraphics[scale=0.3]{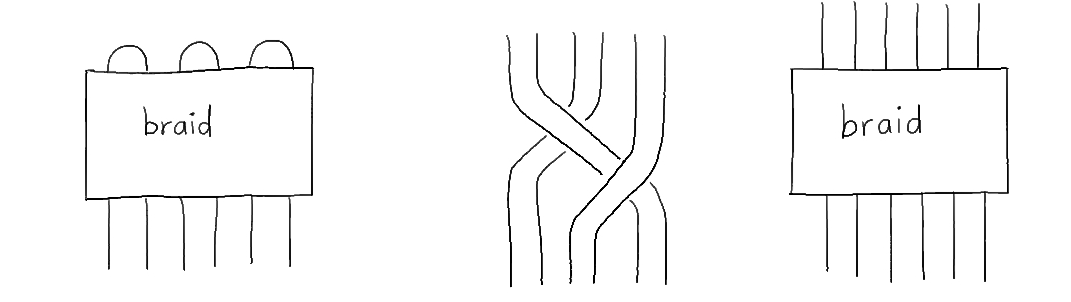}

\caption{Branching sets---\textbf{Left}: 6-Plat for the knot cpomplement, \textbf{Center}: a
3-Braid as 6-Braid as example, \textbf{Right}: 6-Braid for the link complement
\label{fig:Branching-sets-knot-link}}

\end{figure}

Let $C(K)=S^{3}\setminus(K\times D^{2})$ be a knot complement which
is a compact 3-manifold with boundary $\partial C(K)=T^{2}$ the 2-torus.
$C(K)$ is given by a 3-fold branched covering $C(K)\to D^{3}$ inducing
a 3-fold branched covering of the boundary $\partial C(K)\to\partial D^{3}$
or $T^{2}\to S^{2}$. The 3-fold covering $T^{2}\to S^{2}$ has six
points as a branching set, the end points of a 6-plat or tangle (see
Figure \ref{fig:Branching-sets-knot-link}). In a similar manner, let
$C(L)=C(K_{1},K_{2})=S^{3}\setminus\left(L\times D^{2}\right)$ be
a link complement of a link with two linked knots $K_{1},K_{2}$ (the
two linking components). $C(L)$ is a compact 3-manifold with two
boundaries given by two tori. Then, the 3-fold branched covering $C(L)\to S^{3}\setminus\left(D^{3}\sqcup D^{3}\right)$
induces again 3-fold branched coverings $T^{2}\to S^{2}$ of the two
boundaries. Every covering $T^{2}\to S^{2}$ has six points as a branching
set again. The corresponding branching set of $C(L)$ is a braid of
six strands but represented as a 3-braid (see Figure~ \ref{fig:Branching-sets-knot-link}).
Finally, \emph{bosons and fermions are represented as 3-braids.} Our
model agrees with the Bilson--Thompson model but with the exception
that we do not fix the braid. In particular, we do not believe that
the difference between an electron and myon is given by a different
braid.

\section{Electric Charge and Quasimodularity\label{sec:Electric-charge-quasimodularity}}

We argued above that all knot complements admitting a hyperbolic geometry
(geometry of negative, constant scalar curvature) have the properties
of fermions, i.e., spin $\frac{1}{2}$, are pressureless $p=0$ in
cosmology and fulfill the Dirac equation (see also the previous section
for the action functional). However, a particle can carry charges (electrical
or others like color, etc.).

\subsection{Electric Charge as Dehn Twist of the Boundary}

We described above the knot complement as a branched covering branched
along a braid. What is the meaning of a charge in this description?
Let us start with the case of an electric charge. Given a complex line
bundle over $C(K)\times(0,1)$ with connection $A$ and curvature
$F=dA$, we then have the Maxwell equations:
\[
dF=0\qquad d*F=*j,
\]
with the Hodge operator $*$ and the 4-current 1-form $j$. The electric
charge $Q_{e}$ is given by 
\[
Q_{e}=\intop_{\partial C(K)=T^{2}}*F
\]
in the temporal gauge (normal to the boundary of $C(K)$) using $d*F=*j$
and Stokes theorem. The magnetic charge $Q_{m}$ is defined in an
analogous way
\[
Q_{m}=\intop_{\partial C(K)=T^{2}}F=0,
\]
but it is zero because of $dF=0$. By the formulas above, we obtain
a restriction of the complex line bundle to the boundary $\partial C(K)=T^{2}$.
A complex line bundle over $T^{2}$ is determined by the twists of
the fibers w.r.t. the lattice $\mathbb{Z}^{2}$ in the definition
of the torus $T^{2}=\mathbb{C}/\mathbb{Z}^{2}$. However, which twist is
related to the electric charge? Consider a cylinder $S^{1}\times[0,1]$
and identify the ends of the interval to get $T^{2}$ again. A complex
line bundle over $S^{1}\times[0,1]$ with curvature $F$ gives the
integral
\[
\intop_{S^{1}\times[0,1]}F=\intop_{S^{1}\times\left\{ 1\right\} }A-\intop_{S^{1}\times\left\{ 0\right\} }A
\]
using $F=dA$, which is only non-trivial if the two integrals differ.
It can be realized by a twist of one side (say $S^{1}\times\left\{ 1\right\} $),
also called a Dehn twist. Dually by using $d*F=*j$, we get
\[
Q_{e}=\intop_{[0,1]}*j=\intop_{[0,1]}d*F=*F|_{1}-*F|_{0},
\]
which is non-zero by a Dehn twist along $[0,1]$. Therefore, charges
can be detected by Dehn twists along the boundary. A Dehn twist along
the meridian represents the magnetic charge, whereas a Dehn twist along
the longitude is an electric charge. The number of twists is the charge,
i.e., we obtain automatically a quantization of the electric and magnetic
charge. Furthermore, there is a simple algebraic description of the
twists, which agrees with the description of electromagnetic duality
using $SL(2,\mathbb{Z})$. As noted above, the torus can be obtained
by $T^{2}=\mathbb{C}/\mathbb{Z}^{2}$ w.r.t. the lattice $\mathbb{Z}^{2}$.
An~ automorphism of the torus is given by a the group $SL(2,\mathbb{Z})$
acting via rational transformations on $\mathbb{C}$, i.e.,
\[
\left(\begin{array}{cc}
a & b\\
c & d
\end{array}\right)\in SL(2,\mathbb{Z}),\, ad-bc=1\to\left(\begin{array}{cc}
a & b\\
c & d
\end{array}\right)\cdot z=\frac{az+b}{cz+d}.
\]
Then, the two possible Dehn twists are given by
\begin{eqnarray*}
z & \mapsto & z+1,\\
z & \mapsto & -\frac{1}{z},
\end{eqnarray*}
which is known from electromagnetic duality. However, this group also has
another meaning. Let $Diff(T^{2})$ be the diffeomorphism group of
the torus. All coordinate transformations, known as diffeomorphisms
connected to the identity, are forming a (normal) subgroup $Diff_{0}(T^{2})\subset Diff(T^{2})$.
Then, the factor space $MCG(T^{2})=Diff(T^{2})/Diff_{0}(T^{2})$ is
a group, known as a mapping classes group, and generated by Dehn twists,
i.e., $MCG(T^{2})=SL(2,\mathbb{Z})$ or the mapping class group is
the modular group. An element of the mapping class group is a global
diffeomorphism (also called \emph{isotopy}) that cannot be described
by coordinate transformations, i.e., full twists cannot be undone
by a sequence of infinitesimal rotations. Then, different charges belong
to different mapping (or isotopy) classes. Up to now, we have a full
symmetry between electric and magnetic charges (geometrically expressed
by the torus). Now, we will show that this behavior changed for the
extension to the knot complement. Technically, it will be expressed
by a change from modular to quasimodular functions.

\subsection{Electric Charge as a Frame of the Knot Complement}

\begin{figure}
\centering
\includegraphics[scale=0.5]{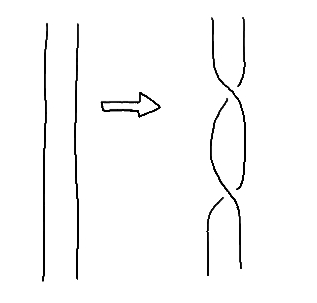}

\caption{Dehn twist represented by a braid.\label{fig:Dehn-twist-represented-braid}}
\end{figure}

However, what does change in the knot complement and in the branched covering?
As a toy example, we consider the complement of the unknot $D^{2}\times S^{1}$.
Then, the Dehn twist along the meridian of the boundary torus will
be trivialized. By a result of McCullough (see for instance \cite{McCulloughHong2013-mcg}),
every Dehn twist along the longitude induces a diffeomorphism of the
solid torus. Then, the complement of the unknot can carry an electric
charge (by a Dehn twist) but no magnetic charge. This result can be
generalized to all knot complements (which are homologically equivalent
to the solid torus). The effect on the branched covering can also be obtained by considering the boundary. The boundary is a torus written
as two-fold branched covering branched along 4 points. A Dehn twist
is given by a permutation of the branching points that leads to a
twist of the braid as a branching set of the knot complement (see Figure
\ref{fig:Dehn-twist-represented-braid}). 
We obtain again the quantization of the electric charge as the number
of Dehn twists.

However, more is true: the isotopy classes of the boundary determine the
isotopy classes of the hyperbolic knot complement up to a finite subgroup
\cite{HatcherMcCullough1997}. The mapping class group $MCG(C(K))$
consists of a disjoint union of isotopy classes of framings, i.e.,
trivializations of the tangent bundle $TC(K)$ seen as sections of
the frame bundle ($SO(3)$ principal bundle) up to homotopy. Therefore,
the change of the number of Dehn twists at the boundary induces a
change of the framing for the knot complement. However, there is also a
direct way using obstruction theory. In \cite{AsselmeyerKrol2009},
we described the quantization of the electric charge by using exotic
smoothness as a substitute for a magnetic monopole. A magnetic monopole
is a substitute for an element in the cohomology $H^{2}(S^{2},\mathbb{Z})$
leading to the quantization of the electric charge
\[
Q_{m}\cdot Q_{e}=\frac{c\hbar}{2}\cdot n,\: n\in\mathbb{Z}
\]
for the magnetic charge $Q_{m}$ and for the electric charge $Q_{e}$.
Using the canonical isomorphism
\[
H^{2}(S^{2},\mathbb{Z})\simeq H^{3}(S^{3},\mathbb{Z}),
\]
we can transform the monopole class (as first Chern class of a complex
line bundle) into a class in $H^{3}(S^{3},\mathbb{Z})$. Now, let $P$
be a principal $SO(3)$ bundle over $S^{3}$, called the frame bundle.
The obstruction for a section in this bundle lies at $H^{4}(S^{3},\pi_{3}(SO(3)))=0$,
where the vanishing of the cocycle guarantees the existence. The number
of sections is given by the elements in $H^{3}(S^{3},\pi_{3}(SO(3)))=H^{3}(S^{3},\mathbb{Z})$
using $\pi_{3}(SO(3))=\mathbb{Z}$. By Hodge duality, we obtain the
same line of argumentation for the class $*F$ getting the electric
charge (using also the quantization condition). The class in $H^{3}(S^{3},\mathbb{Z})$
can be related to a relative class in the 4-manifold $S^{3}\times[0,1]$,
i.e.,
\[
H^{3}(S^{3},\mathbb{Z})\simeq H^{4}(S^{3}\times[0,1],S^{3}\times\partial[0,1],\mathbb{Z})
\]
called the relative Pontrjagin class $p_{1}$. Now, we extend the whole
discussion to an arbitrary 3-manifold $\Sigma$, which we identify
with $\Sigma=C(K)\cup(D^{2}\times S^{1})$. Using this 4-dimensional
interpretation, we obtain the framing as the Hirzebruch defect \cite{Atiyah1990a}.
For that purpose, we consider the 4-manifold $M$ with $\partial M=\Sigma$.
Let $\sigma(M)$ be the signature of $M$, i.e., the number of positive
minus negative eigenvalues of the intersection form of $M$. Furthermore,
let $p_{1}(M,\Sigma)$ be the relative Pontrjagin class as an element
of $H^{4}(M,\partial M=\Sigma,\mathbb{Z})$. Then, the Hirzebruch defect
$h$ is given by
\begin{equation}
h=3\sigma(M)-p_{1}(M,\Sigma)=Q_{e}\label{eq:charge-equals-frame}
\end{equation}
and identified with the framing, i.e., with the charge. This definition
is motivated by the Hirzebruchs signature formula for a closed 4-manifold
relating the signature $\sigma(M)$ and the first Pontrjagin class
$p_{1}(M)$ (of the tangent bundle $TM$) via $\sigma(M)=\frac{1}{3}p_{1}(M)$
(see \cite{Hir:73}).

\subsection{The Charge Spectrum}

Now, we will discuss the expression (\ref{eq:charge-equals-frame})
for the electric charge. By the argumentation above, the relative
Pontrjagin class gives an integer expressing the framing of the knot
complement for a fixed time. The appearance of the signature $\sigma(M)$
added a 4-dimensional element which describes more complex cases with
many components. In \cite{FreGom:91}, this case was also considered
with a similar result: this formula is valid for links where $\sigma(M)$
is now the signature of the linking matrix and $p_{1}$ is the sum
of framings for each component. The signature can be minimally changed
by $\pm1$ leading to a change of the charge by $\pm3$. Therefore,
the minimal change for one component can be
\[
Q_{e}\bmod3\mathbb{Z}=\left\{ 0,\pm1,\pm2,\pm3\right\}, 
\]
i.e., we obtain a spectrum containing four possible values. If we normalize
the charge to be a multiple of $e/3$, then we have the charge spectrum
\[
Q_{e}=\left\{ 0,\pm\frac{1}{3}e,\pm\frac{2}{3}e,\pm e\right\} 
\]
in agreement with the experiment. Then, we have the description of
one particle generation: two leptons (neutrino of charge 0 and lepton
of charge $-1$) and two quarks (quark of charge $-\frac{1}{3}$ and
quark of charge $+\frac{2}{3}$).

\subsection{Vanishing of the Magnetic Charge and Quasimodularity\label{sub:Vanishing-of-magnetic-charge-and-quasimodularity}}

One may wonder whether there is no magnetic charge anymore. Our argument
is only partially satisfying because there are many incompressible surfaces
inside of a hyperbolic knot complement serving as representatives
for magnetic charges. Therefore, we need a stronger argument why the
symmetry between electric and magnetic charge is broken. As explained
above, the Dehn twists of the boundary torus are the generators of
the mapping class (or isotopy) group. According to Atiyah \cite{Atiyah1990a},
the framing can be used to define a central extension $\hat{\Gamma}$
\[
1\to\mathbb{Z}\to\hat{\Gamma}\to\Gamma\to1
\]
of the mapping class group $\Gamma$ so that there is a section $s:\Gamma\to\hat{\Gamma}$
inducing a splitting of the sequence above. This section defines a
canonical 2-cocycle $c$ for the central extension that is given
by the signature of the corresponding 4-manifold (see \cite{Atiyah1990a}
for the details). However, in case of the torus, for the group $\Gamma=SL(2,\mathbb{Z})$,
there is no non-zero homomorphism $\Gamma\to\mathbb{Z}$ and so the
splitting $s_{1}:\Gamma\to\hat{\Gamma}$ is unique. Therefore, the
canonical section $s$ is not a homomorphism and the framing (used
in the definition of this section) leads to a breaking of the modular
invariance i.e., the invariance w.r.t. $\Gamma$. This fact is simply
expressed by considering the difference of the two sections $s(\gamma)$
and $s_{1}(\gamma)$ for $\gamma\in\Gamma$, which is given by the
logarithm of the Dedekind $\eta-$function, related to quasimodular
functions. Thus, our definition of the electric charge breaks explicitly
the electro-magnetic duality and we get a vanishing magnetic charge.

\section{Drinfeld--Turaev Quantization and Quantum States\label{sec:Drinfeld--Turaev-quantization}}

In \cite{AsselmeyerKrol2013,AsselmeyerMaluga2018d}, we discussed the
appearance of quantum states from knots known as Turaev--Drinfeld quantization.
The idea for the following construction can be simply expressed. We
start with two 3-manifolds and consider a cobordism between them.
This cobordism is a 4-manifold with a branched covering branched over
a surface with self-intersections. Here, it is enough to restrict
to a special class of these surfaces, so-called ribbon surfaces (see
\cite{PiergalliniZuddas2005}). The 3-manifolds are chosen to be hyperbolic
knot complements, denoted by $Y_{1},Y_{2}$. A hyperbolic structure
is defined by a homomorphism $\pi_{1}(Y_{i})\to SL(2,\mathbb{C})$
($\in Hom(\pi_{1}(Y_{i}),SL(2,\mathbb{C}))$) up to conjugation. Now,
we extend this structure to the entire cobordism, denoted by $Cob(Y_{1},Y_{2})$.
The branching set of $Cob(Y_{1},Y_{2})$ is a surface $S$ with non-trivial
fundamental group $\pi_{1}(S)$. This surface can be changed without
any change of $Cob(Y_{1},Y_{2})$. One change can be described as
crossing change. Now, we have all ingredients for the Drinfeld--Turaev
quantization:
\begin{itemize}
\item The surface $S$ (branching set of $Cob(Y_{1},Y_{2})$) is inducing
a representation $\pi_{1}(S)\to SL(2,\mathbb{C})$.
\item The space of all representations $X(S,SL(2,\mathbb{C}))=Hom(\pi_{1}(S),SL(2,\mathbb{C}))/SL(2,\mathbb{C})$
has a natural Poisson structure (induced by the bilinear on the group)
and the Poisson algebra \emph{$(X(S,SL(2,\mathbb{C}),\left\{ \,\,\right\} )$}
of complex functions over them is the algebra of observables. 
\item The Skein module $K_{-1}(S\times[0,1])$ (i.e., $t=-1$) has the structure
of an algebra isomorphic to the Poisson algebra $(X(S,SL(2,\mathbb{C})),\left\{ \,\,\right\} )$.\emph{
}(see also \cite{BulPrzy:1999,Bullock1999}).
\item The skein algebra $K_{t}(S\times[0,1])$ is the quantization of the
Poisson algebra $(X(S,SL(2,\mathbb{C})),\left\{ \,\,\right\} )$
with the deformation parameter $t=\exp(h/4)$ (see also \cite{BulPrzy:1999})\emph{
}.
\end{itemize}

\begin{figure}
\centering
\includegraphics[scale=0.25]{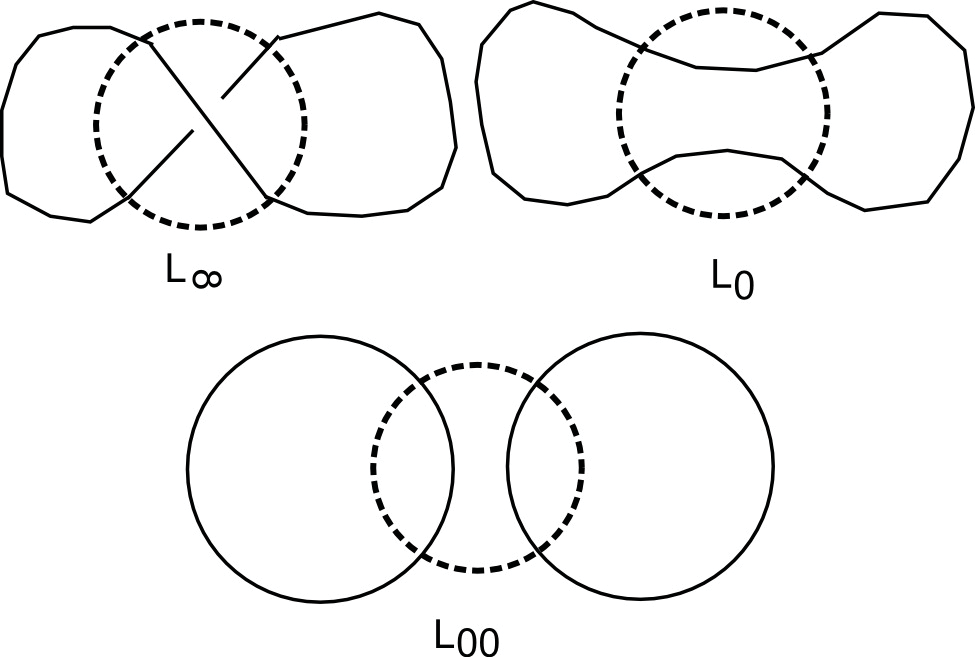}

\caption{Crossings $L_{\infty},L_{o},L_{oo}.$\label{fig:skein-crossings-1}}
\end{figure}

To understand these statements, we have to introduce the skein module
$K_{t}(M)$ of a 3-manifold $M$ (see \cite{PrasSoss:97}). For that
purpose, we consider the set of links $\mathcal{L}(M)$ in $M$ up
to isotopy and construct the vector space $\mathbb{C}\mathcal{L}(M)$
with basis $\mathcal{L}(M)$. Then, one can define $\mathbb{C}\mathcal{L}[[t]]$
as ring of formal polynomials having coefficients in $\mathbb{C}\mathcal{L}(M)$.
Now, we consider the link diagram of a link, i.e., the projection of
the link to the $\mathbb{R}^{2}$ having the crossings in mind. Choose
a disk in $\mathbb{R}^{2}$ so that one crossing is inside this disk.
If the three links differ by the three crossings $L_{oo},L_{o},L_{\infty}$
(see figure \ref{fig:skein-crossings-1}) inside of the disk, then
these links are skein related. 
Then, in $\mathbb{C}\mathcal{L}[[t]]$, one writes the skein relation $L_{\infty}-tL_{o}-t^{-1}L_{oo}$ which depends only on the group $SL(2,\mathbb{C})$, . Furthermore, let $L\sqcup O$ be
the disjoint union of the link with a circle. Then, one writes the framing
relation $L\sqcup O+(t^{2}+t^{-2})L$. Let $S(M)$ be the smallest
submodule of $\mathbb{C}\mathcal{L}[[t]]$ containing both relations.
Then, we define the Kauffman bracket skein module by $K_{t}(M)=\mathbb{C}\mathcal{L}[[t]]/S(M)$.
The modification of $S$ by using the skein relations is one of the
allowed changes of the branching set to keep $Cob(Y_{1},Y_{2})$. 

Now, we list the following general results about this module:
\begin{itemize}
\item The module $K_{-1}(M)$ for $t=-1$ is a commutative algebra.
\item Let $S$ be a surface. Then, $K_{t}(S\times[0,1])$ carries the structure
of an algebra.
\end{itemize}

The algebra structure of $K_{t}(S\times[0,1])$ can be simply seen
by using the diffeomorphism between the sum $S\times[0,1]\cup_{S}S\times[0,1]$
along $S$ and $S\times[0,1]$. Then, the product $ab$ of two elements
$a,b\in K_{t}(S\times[0,1])$ is a link in $S\times[0,1]\cup_{S}S\times[0,1]$
corresponding to a link in $S\times[0,1]$ via the diffeomorphism.
The algebra $K_{t}(S\times[0,1])$ is in general non-commutative for
$t\not=-1$. For the following, we will omit the interval $[0,1]$
and denote the skein algebra by $K_{t}(S)$.

As shown in \cite{AsselmeyerKrol2011d,AsselmeyerKrol2013,AsselmeyerMaluga2016},
the skein algebra serves as the observable algebra of a quantum field
theory. For this approach via branched coverings, the branching sets
of knot complements (representing the fermions) are special braids
(6-plats, see above). Any different braid is a different state or
better than a different quantum state but not a different particle. As
explained above, the charge spectrum is enough to describe one generation
of particles (two leptons, two quarks). The appearance of different
generations will be discussed below.

\section{Fermions and Number Theory}

In this section, we will present some ideas to uncover some explicit
relations between fermions, given as hyperbolic knot complements,
and number theory, notable quaternionic trace fields and algebraic
K theory/Bloch group. However, we first need some definitions.

A quaternion algebra over a field ${\displaystyle F}$ is a four-dimensional
central simple $F$-algebra. A quaternion algebra has a basis ${\displaystyle 1,i,j,ij}$
where ${\displaystyle i^{2},j^{2}\in F^{\times}}$ and ${\displaystyle ij=-ji}$.
A subgroup of ${\displaystyle \mathrm{PSL}(2,\mathbb{C})}$, the isometry
group of the three-dimensional hyperbolic space isomorphic to the
Lorentz group $SO(3,1)$, is said to be derived from a quaternion
algebra if it can be obtained through the following construction.
Let $F$ be a number field that has exactly two embeddings into $\mathbb{C}$
whose image is not contained in $\mathbb{R}$. Let $A$ be a quaternion
algebra over ${\displaystyle F}$ such that, for any embedding ${\displaystyle \tau:F\to\mathbb{R}}$,
the algebra ${\displaystyle A\otimes_{\tau}\mathbb{R}}$ is isomorphic
to the quaternions. Let ${\displaystyle {\mathcal{O}}^{1}}$ be the
group of elements in the order of $A$ with a 1. An order of a quaternionic algebra $A$ is a finitely generated submodule $\mathcal{O}$ of $A$ of reduced norm 1 and let $\Gamma$ be its
image in the $2\times2$ matrices $M_{2}({\mathbb{C}})$ via $\phi:A\to M_{2}(\mathbb{C}).$
We then consider the Kleinian group obtained as the image in${\displaystyle \mathrm{PSL}(2,\mathbb{C})}$
of ${\displaystyle \phi({\mathcal{O}}^{1})}$. This subgroup is called
an arithmetic Kleinian group. An arithmetic hyperbolic three-manifold
is the quotient of hyperbolic space ${\displaystyle \mathbb{H}^{3}}$
by an arithmetic Kleinian group. The complement of the{ figure 8} knot
is one example of an arithmetic hyperbolic 3-manifold.

This class of 3-manifolds shows the strong relation between quaternions
and 3-manifolds. We discussed above the relation between the K3 surface
and the octonions. The starting point for the use of number theory
in Kleinian groups is Mostow's rigidity theorem. A
consequence of this theorem is that the matrix entries in $SL(2,\mathbb{C})$
of a finite covolume Kleinian group $\Gamma$ may be taken to lie
in a number field that is a finite extension of $\mathbb{Q}$. However,
{it is true that} 
there is a strong relation between certain number theoretic
functions (Bloch--Wigner function, dilogarithm) and the volume of the
hyperbolic 3-manifolds: the volume is the sum over all Bloch--Wigner
functions for the ideal tetrahedrons forming this 3-manifold. For
more information about the relation between hyperbolic 3-manifolds
and number theory, consult the book \cite{MacLachlanReid}. We hope
to use this relation in the future to obtain more properties of fermions
by using number theory.

\section{The K3 Surface and the Number of Generations}

In Section \ref{sub:K3-surfaces-and-octonions}, we discussed a
relation between the K3 surface and octonions by using the intersection
form. Here, we use only the $E_{8}$ matrix, i.e., the Cartan matrix
of the Lie algebra $E_{8}$. In this section, we will speculate about
the other part
\[
H=\left(\begin{array}{cc}
0 & -1\\
-1 & 0
\end{array}\right)
\]
of the intersection form (now with a different orientation). $H$
is the intersection form of the 4-manifold $S^{2}\times S^{2}$. To
express it explicitly, there are homology classes $\alpha,\beta\in H_{2}(S^{2}\times S^{2})$
with $\alpha^{2}=\beta^{2}=0$ and $\alpha\cdot\beta=-1$. Therefore,
every $S^{2}$ of this manifold has no self-intersections. For the
topology of the K3 surface with intersection form (\ref{eq:intersection-K3}),
this form has the desired form, but, as explained above, we will change
the smoothness structure. The central idea is the usage of Casson
handles $CH$ for the 4-manifolds $S^{2}\times S^{2}\setminus pt$,
the one-point complement of $S^{2}\times S^{2}$. Here, the homology
classes $\alpha,\beta$ are given (up to homotopy) by $\alpha^{2},\beta^{2}=0\bmod2$
and $\alpha\cdot\beta=-1$; see \cite{Cas:73}. However, then one has $\alpha^{2}=2n$.
Interestingly, the existence of a spin structure is connected to the
property that the squares of the homology classes are even. Here, we
will consider the simplest realization which has non-zero squares,
i.e., we get the form
\begin{equation}
\tilde{H}=\left(\begin{array}{cc}
2 & -1\\
-1 & 2
\end{array}\right).\label{eq:modified-H}
\end{equation}
This form cannot be an intersection form because $\tilde{H}$ has
determinant $3$. Therefore, only the $\tilde{H}\bmod2$ reduction
has the meaning to be an intersection. However, for the moment, we will
consider $\tilde{H}$ and apply the same construction as for the $E_{8}$,
i.e., we see $\tilde{H}$ as the Cartan matrix for a Lie algebra. In
this case, we get the Lie algebra of $SU(3)$ or the color group.
The whole discussion uses some hand-waving arguments, but it is a sign
that the $3\left(S^{2}\times S^{2}\right)$ part of the K3 surfaces
is connected with the generations. Every part $S^{2}\times S^{2}$
has one color group and realizes the electric charge spectrum $0,\pm\frac{1}{3},\pm\frac{2}{3},\pm1$.
Thus,  every $S^{2}\times S^{2}$ is the 4-dimensional expression for
one generation. This result agrees with the discussion in \cite{AsselmeyerBrans2015}
where we generate fermions from a Casson handle. Let us assume that
the number of generations is given by the number of $S^{2}\times S^{2}$
summands. How many generations are possible? Here, we have the surprising
result: \emph{if the underlying spacetime is a smooth manifold, then
the minimal number of generations must be three!} A spacetime with
only one or two generations is not a smooth manifold. Then, the K3
surface is the minimal model.

We will close this paper with another speculation, a global symmetry
induced from the K3 surface. Starting point is the intersection form
again. From the point of number theory, this form is an even unimodular
positive-definite lattices of rank 24, the so-called Niemeier lattice.
In 2010, Eguchi--Ooguri--Tachikawa observed that the elliptic genus
of the K3 surface decomposes into irreducible characters of the N
= 4 superconformal algebra. The corresponding q-series is a mock modular
form related to the sporadic group $M_{24}$, the Mathieu group, a
simple group of order 244823040. The whole theory is known as Mathieu
moonshine or umbral moonshine \cite{umbral-moonshine}. The interesting
point here is the maximal subgroup of $M_{24}$, the split extension
of $PGL(3,4)$ by $S_{3}$. The group is the projective group of $3\times3$
matrices with values in the field of four elements. It seems that this
maximal subgroup acts in some sense on the K3 surface, and we conjecture
that this group acts on the $S^{2}\times S^{2}$ part. If our idea
of a relation between the three generations and the $3\left(S^{2}\times S^{2}\right)$
part of the K3 surfaces is true, then we hope to get the mixing matrix
for quarks and neutrinos from this action.

\section{Conclusions and Outlook}

In this paper, we presented a top-down approach to fermions and bosons,
in particular the standard model. What was done in the paper?
\begin{itemize}
\item We constructed a spacetime, the K3 surface and derive some numbers
like the cosmological constant or some energy scales and neutrino
masses agreeing with experimental data.
\item We derived from a representation of K3 surfaces by branched covering
a simple picture: fermions are hyperbolic knot complements, whereas
bosons are link complements (torus bundles).
\item We obtained the gauge group from this picture (at least in principle).
\item We derived the correct charge spectrum and obtained one generation.
\item We conjectured about the number of generations and global symmetry
(the $PGL(3,4)$) to get the mixing between the generations.
\end{itemize}

{ 
What are the consequences for physics? The model only has a few direct consequences. We introduced fermions and bosons in a geometric way. Except for the right-handed neutrino (needed for the see-saw mechanism to generate the masses), we only got the fermions and gauge bosons of the standard model. No extension is needed. The usage of torus bundles for the gauge bosons should generate additional relations for the corresponding scattering amplitudes. The appearance of the global symmetry $PGL(3,4)$ should be related the mixing of quarks and neutrinos. In \cite{AsselmeyerKrol2018b}, we also discussed the appearance of an asymmetry between particles and anti-particles induced by the topology of the spacetime. This idea is also valid in this model, but we cannot match it to the observations.

Is there an outline on some new experiments derived by this model? Currently, this model makes some predictions about the neutrino masses, charge spectrum and the existence of a right-handed neutrino. However, these predictions can be checked by a better measurement in known experiments. Now, there are no new ideas about special experiments connected with this model.
}

Among these results, there are, of course, many open points of the kind:
what is the color and weak charge? How can we implement the Higgs
mechanism? What is mass? For the Higgs mechanism, we had found a possible
scheme in our previous work \cite{AsselmeyerKrol2013b,AsselmeyerKrol2018b},
but it is only a beginning. Many aspects of this paper are related
to the ideas of Furey and Gresnigt. It is a future project to extend it
and bridge our approach with these ideas.

\vspace{12pt}
\section*{Acknowledgments}
{I first want to thank the anonymous referees for all helpful remarks and comments to make this work more readable. Furthermore, I want to thank Carl Brans for an uncountable number of delightful
discussions over so many years. I also want to give a special thanks to Chris Duston for
discussions about branched coverings, in particular to draw my attention
towards branched coverings. In addition, I want to acknowledge all of my discussions
with Jerzy Krol over the years. Finally, I want to give a huge thank you to my family, in particular to my daughter Lucia, for painting
the figures.} 

\end{document}